\documentclass[a4paper,twocolumn,prb]{revtex4}
\usepackage{amsmath}
\usepackage{graphicx}
\usepackage{color}
\usepackage{subfigure}
\usepackage{hyperref}
\begin{document}
\title{Numerical Determination of Boundary Condition Changing Operators}
\author {M. N. Najafi }
\email{morteza.nattagh@gmail.com}
\affiliation{Physics department, Sharif University of Technology, P.O. Box 11155-9161, Tehran, Iran}

\begin{abstract}
A new numerical method to determine the boundary condition changing (bcc) operators in the statistical models is introduced. This method is based on a variant of Schramm-Loewner Evolution (SLE), namely SLE($\kappa,\rho$). As a prototype, Abelian Sandpile Model (ASM) with a sink on some point on the boundary is considered. Using this method we study the bcc operator corresponding to sink point on the boundary. It is numerically shown that the conformal dimension of the this operator is nearly $0$. The most appropriate candidate for this operator is the logarithmic partner of the unity operator $\tilde{I}\equiv :\theta\bar{\theta}:$, as it has been conjectured theoretically.
\end{abstract}
\maketitle
\section{Introduction}

   In determining the operator content of boundary conformal field theory (BCFT), one of the most important challenges is to obtain the boundary condition changing (bcc) operators corresponding to the changes of the boundary conditions. This question has been addressed in some minimal models, such as percolation \cite{Cardy1} and Ising model with spin change on the boundary \cite{Ising} etc. To determine the conformal weight of these bcc operators, one has to determine the correlation functions of the model and compare them with the exact results, or use some evidences e.g. operator product expansion (OPE) of these operators \cite{francesco}. This needs almost complete knowledge of the operator content of the model in hand which for the less-known models is not applicable. Therefore one needs more direct method to study such operators. An important recent discovery, concerning this difficulty, was the discovery of the bcc operator corresponding to the change of boundary conditions of the points at which the trace of a critical interface starts. These interfaces are some non-intersecting macroscopic objects which separate the different phases of the model and are postulated to be described by Schramm-Loewner Evolution (SLE) theory whose operator content is almost well understood\cite{Shramm}.
   
   Schramm-Loewner evolution, as the new interpretation of statistical models, has attracted much attention in last years. According to this theory one can classify 2D statistical models via a growth processes by focusing on their geometrical objects, such as interfaces. The essential building block of SLE is the conformal symmetry of probability measures of the model in hand. Therefore it seems natural to exist a connection between SLE and conformal field theory (CFT) in which, contrary to SLE, one deals with the local fields. This relation shows itself in a simple relation between the central charge $c$ in CFT and the diffusivity parameter $\kappa$ in SLE \cite{BauBer}. The parameter $\kappa$ is a quantity which represents the universality class in which the statistical models belong to. Due to its capability in reflecting essential properties of the models, SLE has been referred
and employed theoretically and numerically in various statistical models such as turbulence \cite{BerBofCelFal}, ASM avalanche frontier \cite{Saberi1}, iso height lines of KPZ \cite{Saberi2}, WO3 \cite{Saberi3}, Ising model \cite{Saberi4}, etc. to study statistical properties of them. Equally much theoretical works have been carried out on this theory, such as left passage probability and fractal dimension of the curves \cite{cardy}, crossing probability \cite{Cardy1}, etc.

In this paper we introduce a new method to determine numerically the conformal weight of the bcc operator using a generalization of SLE. This generalization, namely SLE($\kappa,\rho$) governs the law of the curves whose growth media contains, in addition to the starting and ending points of the curve, some other preferred points. In this theory $\rho$ has to do with the conformal weight of the corresponding bcc operator and has the capability to yield some informations about this operator.  We test this method on the Abelian sandpile model (ASM) to determine the conformal weight of the sink points (at which the grains of the sand pile dissipate) on the boundary of the domain in which the model is defined. Although there are some theoretical evidences that in this model, the bcc operator is the logarithmic partner of the identity operator i. e. $\tilde{I}\equiv :\theta\bar{\theta}:(z)$, since ASM is somehow problematic, we believe that it is interesting to study and investigate this operator directly using our method.

The next section is devoted to introduction to SLE and its generalization SLE($\kappa,\rho$). In section \ref{ASM} we introduce in sum the ASM and define our main concern in this paper and in section \ref{num} we present the numerical results of the application of our method mentioned above.

\section{SLE}\label{SLE}
In two dimensions, the phase boundaries of the statistical models form some non-intersecting random curves which directly reflect the status of the system in question. This theory is applicable to the curves which are supposed to have two properties: conformal invariance and the domain Markov property. According to this theory, each conformal invariant statistical model fall into a one-parameter universality class which is represented by $\kappa$. We offer a very brief introduction below; for good introductory reviews see references \cite{RohdeSchramm,cardy}.
\subsection{Chordal SLE}
Let us denote the upper half-plane by $H$ and $\gamma$ as the SLE trace. Then parametrize the curve with "time" $t$ and name the curve up to this time as $\gamma_t$. Define $\tau_z$ as the time at which this curve meets the point $z$ in the upper half plane. The hull of this growth process is defined as $K_{t}=\overline{\lbrace z\in H:\tau_{z}\leq t \rbrace}$ so that the complement $H_{t}:=H\backslash{K_{t}}$ is simply-connected. According to Riemann's mapping theorem there is always a unique conformal map $g_{t}(z)$ which maps $H_{t}\rightarrow{H}$ with $g_{t}(z)=z+\frac{2t}{z}+O(\frac{1}{z^{2}})$ as $z\rightarrow{\infty}$ known as hydrodynamical normalization. SLE$_{\kappa}$ is then defined via such conformal maps which are solutions of Loewner's equation:
\begin{equation}
\partial_{t}g_{t}(z)=\frac{2}{g_{t}(z)-\xi_{t}},
\label{Loewner}
\end{equation}
with the initial condition $g_{t}(z)=z$. In this equation the driving function $\xi_{t}$ is proportional to a one dimensional standard Brownian motion i.e. $\xi_{t}=\sqrt{\kappa}B_{t}$. For fixed $z$, $g_{t}(z)$ is well-defined up to time $\tau_{z}$ for which $g_{t}(z)=\xi_{t}$. One can retrieve the SLE trace by $\gamma_{t}=\lim_{\epsilon\downarrow{0}}g_{t}^{-1}(\xi_{t}+i\epsilon)$. There are phases for these curves, for $0<\kappa\leq{4}$ the trace is non-self-intersecting and it does not hit the real axis;  in this case the hull and the trace are identical: $K_{t}=\gamma_{t}$. This is called "dilute phase". For $4<\kappa<{8}$, the trace touches itself and the real axis so that a typical point is surely swallowed as $t\rightarrow\infty$ and $K_{t}\neq\gamma_{t}$. This phase is called "dense phase". And for $\kappa\geq 8$ the curve is space-filling. The frontier of $K_{t}$, i.e. the boundary of $H_{t}$ minus any portions of the real axis for $4<\kappa<8$ is a simple curve which is locally a SLE$_{\tilde{\kappa}}$ curve with $\tilde{\kappa}=\frac{16}{\kappa}$ ($2<\tilde{\kappa}<4$), i.e. it is in the dilute phase \cite{Dub}. This relation relates two phases to each other.

\subsection{SLE($\kappa,\rho$)}\label{SLE(k,r)}
SLE theory describes the critical random curves which, except the origin (at which the curve starts to grow) and the ending point (which is infinity in the chordal case), there is no other preferred points on the real axis. But in some situations the existence of some preferred points on the real axis does affect the growth of the curve. As an example let us consider a curve that starts from the origin and ends on a point on the real-axis ($x_{\infty}$). In this situation, we will have two preferred points on the real axis. Using the conformal map $\phi={x_{\infty}z}/{(x_{\infty}-z)}$, one can send the end point of the curve to the infinity. In this respect, the function $h_{t}=\phi{\circ }g_{t}{\circ}\phi^{-1}$ describes chordal SLE. It is easy to show that the equation governing $h_{t}$ is $\partial_{t}g_{t}=2/(\lbrace{\phi'(g_{t})(\phi(g_{t})-\xi_{t})}\rbrace)$. But it is explicit that this function is not hydrodynamically normalized. It has been shown \cite{BauBer2} that if one uses another mapping $\tilde{g}_{t}=v_{t}{o}h_{t}{o}u^{-1}$ where $u=\phi^{-1}$ and $v_{t}$ is a linear fractional transformation that make the corresponding map hydrodynamically normalized, then the stochastic equation of $\tilde{g}_{t}$ is the same as Eq. (\ref{Loewner}). In fact this procedure leaves the Eq. (\ref{Loewner}) unchanged but leads the driving function to have a drift term \cite{BauBer2}:
\begin{equation}
d\xi_{t}=\sqrt{\kappa}dB_{t}+\frac{\kappa-6}{\xi_{t}-g_{t}(x_{\infty})}dt
\label{driving}
\end{equation}
Thus for the critical curves from boundary to boundary, the corresponding driving function acquires a drift term. This generalization of SLE can be generalized further to have multiple preferred real axis points. For review see references \cite{RohdeSchramm,cardy}. Eq[\ref{driving}] is especial case of the more general theory i.e. SLE($\kappa,\rho=\kappa-6$). \\
In the SLE($\kappa,\rho$), the parameter $\kappa$ identifies the local properties of the model in hand and corresponds directly to the central charge of the corresponding conformal field theory and the parameter $\rho$ has to do with the boundary conditions (bc) imposed i.e. some information of the bcc operator have been coded in this parameter. The actual behavior depends on the concrete values of $\kappa$ and $\rho$. The example is the dipolar SLE($\kappa$) in which $\rho=(\kappa-6)/2$ \cite{BerBau}. The other example is the situation in which except the origin and the ending point (say infinity), the boundary condition changes at the other point $x_0$. The stochastic equation governing such curves is the same as formula (\ref{Loewner}) but the driving function has a different form: 
\begin{equation}
d\xi_{t}=\sqrt{\kappa}dB_{t}+\frac{\rho}{\xi_{t}-g_{t}(x_0)}dt
\label{SLE(k,r)driving}
\end{equation}
The generalization of Eq [\ref{SLE(k,r)driving}] is direct for more preferred points on the real axis ($x_0,x_1,...,x_n$) \cite{Kytola}:

\begin{equation}
d\xi_{t}=\sqrt{\kappa}dB_{t}+\sum_{i=0}^{n}\frac{\rho_i}{\xi_{t}-g_{t}(x_i)}dt
\label{SLE(k,r)driving-general}
\end{equation}

\section{Introduction to ASM and the Problem Definition}\label{ASM}
The simple model of Bak, Tang and Wiesenfeld \cite{Bak} of Self-Organized Criticality (SOC) phenomena has attracted much attention due to its rich structure and complex behaviors. As an example of this phenomena, they introduced the sandpile models in which without tuning any parameter, the system show critical (such as power law) behaviors. The abelian structure of this model was first discovered by D. Dhar and named as Abelian Sandpile Model (ASM) \cite{Dhar1}.  ASM has various and interesting features and many different analytical and numerical works have been done on this model. For example different height and cluster probabilities \cite{Majumdar3}, its connection with spanning trees \cite{Majumdar2}, ghost models \cite{Mahieu}, q-state Potts model \cite{Saluer}, etc. For a good review see reference \cite{Dhar2}.

\begin{figure}
\centerline{\includegraphics[scale=.40]{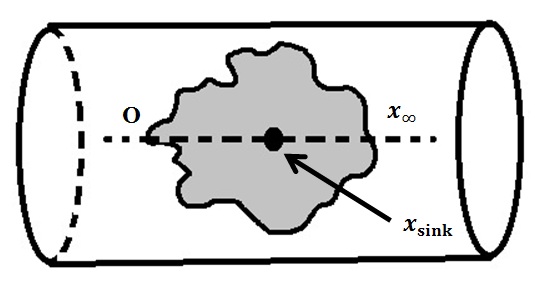}}
\caption{The schematic picture of the Abelian sandpile model on the cylinder in which the grey region has toppled.}
\label{sample}
\end{figure}

Before going to a specific geometry, let us consider the ASM on a two-dimensional square lattice $L\times{L}$. To each site $i$, a height variable $h_{i}$ is assigned taking its values from the set {1,2,3,4} the number of sand grains on this site. Define the dynamics of this model as follows; in each step, a grain is added to a random site $i$ i.e. $h_{i}\rightarrow{h_{i}+1}$; if the resulting height becomes more than 4, the site topples and loses 4 sand grains, each of which is transfered to one of the four neighbors of the original site. As a result, the neighboring sites may become unstable and topple and a chain of topplings may happen in the system. At the boundary sites, the toppling causes one or two sand grains leave the system. This process continues until the system reaches a stable configuration. Now another random site is selected and the sand is released on this site and the process continues. After a finite number of steps, the ASM reaches a well-defined distribution of states in which some configurations do not occur (transient configurations)  and other configurations occur with equal probability (recurrent configurations). For details see reference \cite{Dhar2}. This model is related to Potts model with $q\rightarrow 0$ and CFT with central charge $c=-2$. For a lattice with $d$ neighboring sites, the toppling occurs when $h_{i}>d$, then the original site will lose $d$ grains and the height of each of its neighbors will increase by 1. It has been shown that for such a system, the avalanche frontier is a random loop with the fractal dimension $\frac{5}{4}$ and the same statistical properties as the loop erased random walk (LERW)\cite{najafi2}. To determine the diffusivity parameter $\kappa$ we should find some random curves which goes to a point at infinity whereas the mentioned frontier forms a closed loop. To make such curves, one can cut the loops horizontally and send the end point of the curve to the infinity. This method causes some large numerical errors. It has been proposed in \cite{najafi1} that one can skip the mapping to infinity and interpret the resulting curve as a curve which start and end on the real axis. In this situation the formalism of SLE($\kappa,\rho$) can be applied to find the diffusivity $\kappa$ and $\rho$ parameters of the model. To this end the uniformizing map firstly should become disceretized $G_{t_n,\xi_n}=G_{\delta t_{n-1},\xi_{n-1}}{o}G_{\delta t_{n-2},\xi_{n-2}}{o}...{o}G_{\delta t_{0},\xi_{0}}$, ($\delta t_{n}=t_{n}-t_{n-1}$) in which 
\begin{equation}
G_{\delta t_{n},\xi_{n}}(z)=\xi_{n}+\sqrt{(z-\xi_{n})^{2}+4\delta t_{n}}
\label{slit}
\end{equation}   
 and then assuming the driving function partially constant (in each time interval), so the Eq [\ref{driving}] can become discretized as follows:
\begin{equation}
\delta\xi_{n}=\sqrt{\kappa}\delta{B_{n}}+\frac{\rho}{\xi_{n}-G_{t_{n}}(x_{\infty})}\delta{t_{n}}.
\label{dis driving}
\end{equation}
Rewriting this equation in the form:
\begin{equation}
\frac{\xi_{n}-\sum_{i=1}^{n}[\frac{\rho\delta{t_{i}}}{\xi_{i}-G_{t_{i}}(x_{\infty})}]}{\sqrt{\kappa}}=B_{n}.
\label{BM}
\end{equation}
one can find the corresponding $\kappa$ and $\rho$ by demanding that the right hand side of Eq [\ref{BM}] be a one dimensional Brownian motion. It has been shown that this method results in more precise and reliable determination of these parameters \cite{najafi1}.\\
Now consider one additional preferred point on the real axis sitting in $x_0$. The random curves in this media are described by SLE($\kappa,\kappa-6,\rho$) in which the parameter $\rho$ is related to the bcc operator in $x_0$. Then the Eq [\ref{SLE(k,r)driving-general}] yields:
\begin{equation}
d\xi_{t}=\sqrt{\kappa}dB_{t}+\frac{\kappa-6}{\xi_{t}-g_{t}(x_{\infty})}dt+\frac{\rho}{\xi_{t}-g_{t}(x_0)}dt.
\label{driving2}
\end{equation}
There is a simple relation between the conformal weight of bcc operator at $x_0$, $h_{\rho}$ and $\rho$ i.e. $h_{\rho}=\frac{\rho(\rho+4-\kappa)}{4\kappa}$. All the arguments mentioned above are applicable to this case i.e. one can determine the unknown parameters of the model by fitting the following equation by one dimensional Brownian motion:
\begin{equation}
\frac{\xi_{n}-\sum_{i=1}^{n}[\frac{(\rho_c)\delta{t_{i}}}{\xi_{i}-G_{t_{i}}(x_{\infty})}]-\sum_{i=1}^{n}[\frac{\rho\delta{t_{i}}}{\xi_{i}-G_{t_{i}}(x_0)}]}{\sqrt{\kappa}}=B_{n}.
\label{BM2}
\end{equation}

The best value of $\rho$ is such a parameter which results in a nice one-dimensional Brownian motion in right hand of Eq [\ref{BM2}]. Since the numerical calculation of these three parameters togather is difficult, the best method to obtain this quantity is to obtain best values of $\kappa$ and $\rho$ for the case there is no boundary condition change, and then repeat the calculations for the case with boundary condition change, setting $\kappa$ and $\rho_c$ fixed. \\

\begin{figure}
\centerline{\includegraphics[scale=.55]{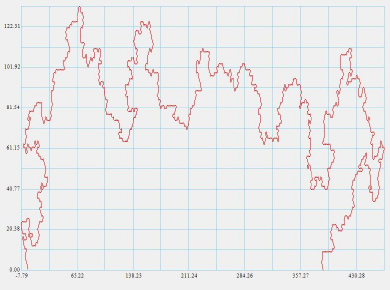}}
\caption{A sample curve which is the cut exterior boundary of an avalanche.}
\label{sample1}
\end{figure}

\section{Numerical Methods and Results}\label{num}
Consider ASM on a cylinder with length and circumference $L$ in presence of a sink point in which the grains dissipate (the point $x_0$). Fig [\ref{sample}] shows schematically such a model in which the grey area is the set of toppled sites. Therefore in one direction the boundary conditions are periodic in contrast to the other direction in which the sand grains may leave the system i.e. with open boundary condition. This geometry is compatible with our purpose i.e. we have real boundary at which the sink point has been located. Now we consider the upper portion of the loop (the frontier of the avalanche) in such a way that the resulting curve goes from origin to the point $x_{\infty}$ on the boundary in presence of the the sink point at $x_0$. If $L$ be much larger than the linear size of the curves, we can approximately ignore the finite size effect and apply the chordal SLE formalism. We also rescale the curves in such a way that $x_0-\xi_0=1$. In figure \ref{sample1} we have presented a sample curve which is the cut exterior boundary of an avalanche. In \cite{Saberi1} by sending the end point of such a curve to infinity using making chordal SLE, it was shown that the amount of $\kappa$ for this problem is nearly $2$ (without dissipation in sink). In this paper we do not apply such a infinity map due to the reasons to be mentioned in the following subsections.\\
The simulation has been done over $3\times 10^4$ samples at which the minimum size of the curves is $1000$ in lattice units. The CPU  time  for this simulation was nearly $3\times 10^6$s for each case\footnote{We have used Intel dual core CPU (3.2 GHz) desktop computer.}. 

\begin{figure}
\centerline{\includegraphics[scale=.36]{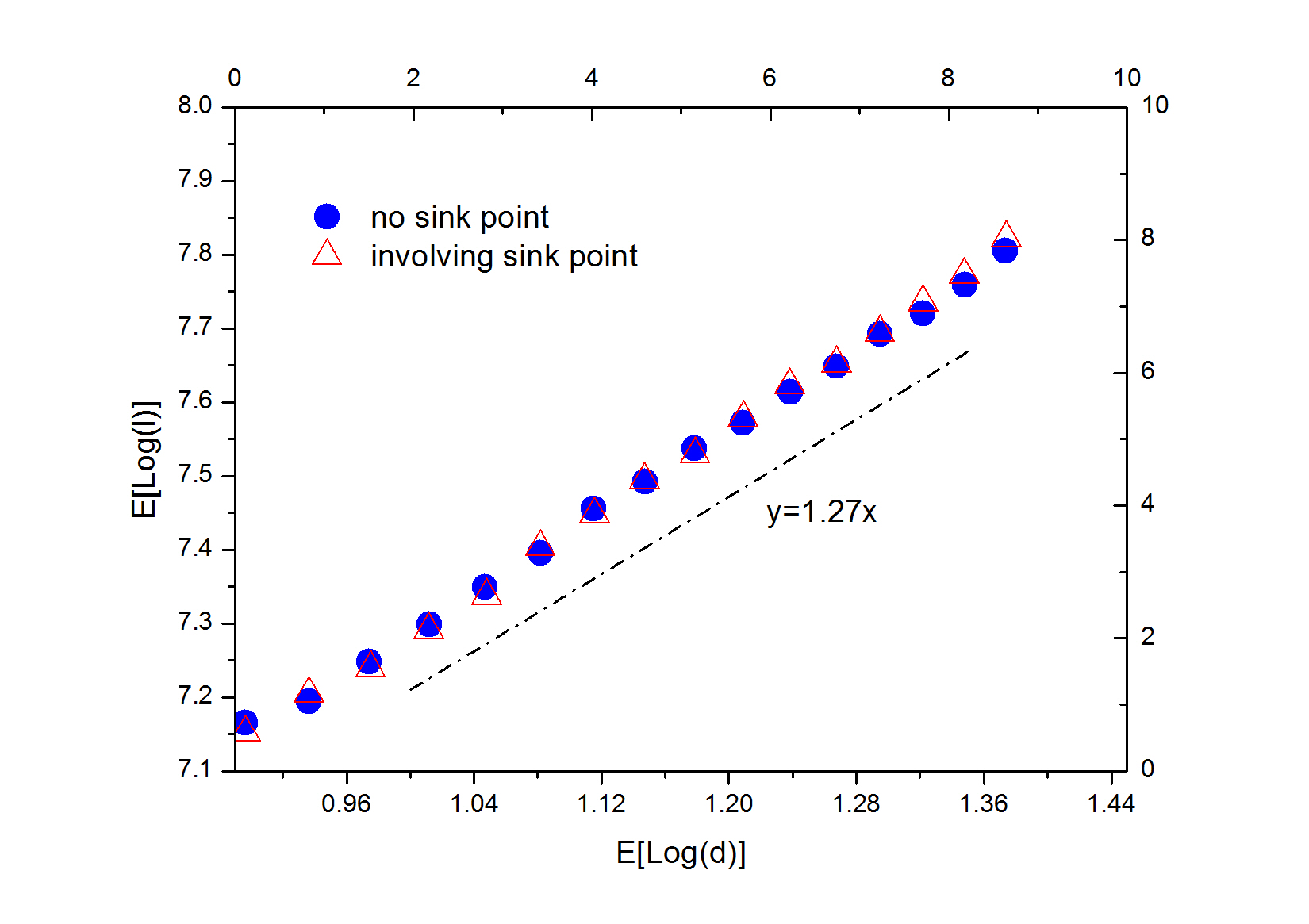}}
\caption{The fractal dimension of the curves in two cases; with and without sink point.}
\label{FrDim}
\end{figure}

\begin{figure}
\centerline{\includegraphics[scale=.36]{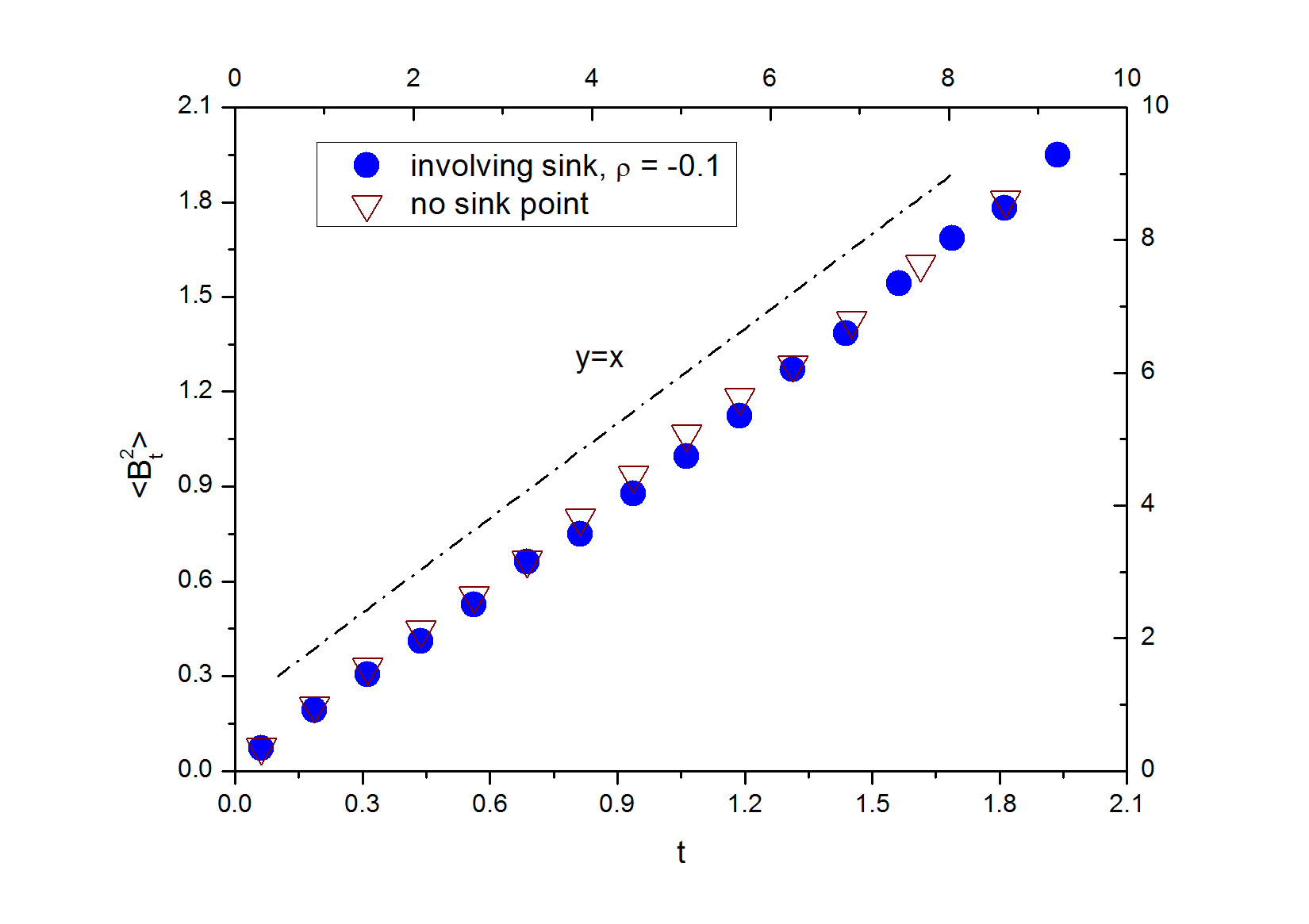}}
\caption{The graph of $\left\langle B_t^2\right\rangle$ defined in Eq[\ref{BM2}] versus $t$ for the ASM without sink point ($\kappa=2.0$ and $\rho_c=-3.2$) and the case involving sink point with the same $\kappa$ and $\rho_c$ and $\rho=-0.1$.}
\label{BrMo}
\end{figure}

\subsection{Fractal dimension}\label{fracdim} To avoid the possible errors due to existence of couple of equations in the SLE($\kappa,\rho_c,\rho$) formalism, we have made two parallel simulations with the same conditions, one for the case in which there is no sink point, and another for the case involving the sink point. One can calculate the amount of $\rho$ by comparing the resulting Brownian motion of the two simulations. First of all, we should determine the best values for $\kappa$ and $\rho_c$ for the first case and then fixing these quantities, we can find the best value of $\rho$ for the second case for which the two simulations best coincide. To this end we have calculated the fractal dimension of the curves in each case. We use the fact that (showing the linear size of the curve by $d$ and the length of the curve by $l$) $l\sim d^{D_f}$ where $D_f$ is the fractal dimension of the curve. The Fig [\ref{FrDim}] in which the log-log plot of $l$ and $d$ is sketched, demonstrates that the fractal dimension of the curves in two cases coincides and is nearly $\frac{5}{4}$. The main result in this subsection is that the amount of $\kappa$ and $\rho_c$ in the two cases is equal since the fractal dimension is only related to the amount of $\kappa$ i.e. $D_f=1+\frac{\kappa}{8}$ and $\rho_c$ only depends on $\kappa$ \cite{cardy}.

\subsection{Calculation of $\rho$} To use the result of the subsection \ref{fracdim}, we should obtain $\kappa$ and $\rho_c$ for the case without sink point, and then use these parameters for calculationg $\rho$ for the case involving sink point. The first simulation has been done in \cite{najafi1} in which it was shown that the best values are $\kappa=1.95\pm 0.07$ and $\rho_c=3.5\pm 0.5$. Our simulation for our case (which is on cylinder and the curves are rescaled) coincides with this report, i.e. we have obtained $\kappa=2.0\pm 0.1$ and $\rho_c=3.2\pm 0.5$. The most important part of our results is the second part of simulation, i.e. determination of $\rho$. Fig [\ref{BrMo}] shows the variance of Brownian motion for the two mentioned cases. We see that the two curves coincide by attributing to $\rho$ the value $-0.1$. By using Maximum Likelihood Estimation (MLE) \cite{MLE} we have obtained $\rho=-0.1\pm 0.2$ with the probability $0.8$. From this result one can calculate the conformal weight of the bcc operator corresponding to sink; $h_{\rho}=0.0_{-0.05}^{+0.02}$.

Our result is in agreement with the prediction of field theoretical point of view of ASM. It has been conjectured that the sink point, which can be interpreted as a change of boundary conditions from closed to open and immediately from open to closed, is equivalent to putting an operator resulting from the OPE of two twist operators $\mu$ with the conformal weight $-\frac{1}{8}$. According to \cite{Ruelle}, the OPE of $\mu$ operator is the direct sum of identity operator and its logarithmic partner and the second one corresponds to the sink point which is our case. So the result of this OPE is nothing but the logarithmic partner of identity operator $\tilde{I}=:\theta\bar{\theta}:(z)$ with the conformal weight $h_{\rho}=0$. 

\section{Conclusion}
In this paper we have proposed a framework to obtain numerically the conformal weight of the boundary condition changing (bcc) operator in a critical statistical model. As an example, Abelian sandpile model (ASM) has been considered. Specially we have considered a sink point on the boundary in which the grains dissipate and have calculated the conformal weight corresponding to the bcc operator to be $h_{\rho}=0.0_{-0.05}^{+0.02}$ which is in agreement with the field theoretical result $h_{\rho}=0$ corresponding to the conformal weight of logarithmic partner of the identity operator $\tilde{I}=:\theta\bar{\theta}:(z)$.

\end{document}